\documentclass[twocolumn,prl,showpacs,nofootinbib,superscriptaddress]{revtex4}

\usepackage{graphicx}
\usepackage{amsmath,amsfonts,amssymb,bm}

\usepackage[usenames]{color}

\begin{document}

\vspace*{0mm}
\title{Muonic hydrogen cascade time and lifetime of the short-lived 
       $2S$ state}

\author{L.~Ludhova}
\altaffiliation[Present address: ]{Dipartimento di Fisica, 
               Universit\`a degli Studi, via Celoria 16, 20133 Milano, Italy}
\altaffiliation[Electronic address: ]{Livia.Ludhova@mi.infn.it}
\affiliation{D\'epartement de Physique, Universit\'e de Fribourg,
             1700 Fribourg, Switzerland }
\affiliation{Paul Scherrer Institut, 5232 Villigen-PSI, Switzerland }
\author{F.D.~Amaro}
\affiliation{Departamento de F{\'\i}sica, Universidade de Coimbra, 
             3000 Coimbra, Portugal }
\author{A.~Antognini}
\affiliation{Max-Planck-Institut f\"ur Quantenoptik, 85748 Garching, 
             Germany}
\author{F.~Biraben}
\affiliation{Laboratoire Kastler Brossel, ENS, UPMC and CNRS,
             4 place Jussieu, 
             75252 Paris Cedex 05, France }
\author{J.M.R.~Cardoso}
\affiliation{Departamento de F{\'\i}sica, Universidade de Coimbra, 
             3000 Coimbra, Portugal  }
\author{C.A.N.~Conde}
\affiliation{Departamento de F{\'\i}sica, Universidade de Coimbra, 
             3000 Coimbra, Portugal  }
\author{A.~Dax}
\altaffiliation[Present address: ]{Department of Physics,
             University of Tokyo, 7-3-1 Hongo, Bunkyo-ku,
             Tokyo 113-0033, Japan}
\affiliation{Physics Department, Yale University, New Haven, CT 06520-8121, 
             USA }
\affiliation{Paul Scherrer Institut, 5232 Villigen-PSI, Switzerland }
\author{S.~Dhawan}
\affiliation{Physics Department, Yale University, New Haven, CT 06520-8121, 
             USA }
\author{L.M.P.~Fernandes}
\affiliation{Departamento de F{\'\i}sica, Universidade de Coimbra, 
             3000 Coimbra, Portugal  }
\author{T.W.~H\"ansch}
\affiliation{Max-Planck-Institut f\"ur Quantenoptik, 85748 Garching, 
             Germany}
\author{V.W.~Hughes}
\altaffiliation{Deceased}
\affiliation{Physics Department, Yale University, New Haven, CT 06520-8121, 
             USA }
\author{P.~Indelicato}
\affiliation{Laboratoire Kastler Brossel, ENS, UPMC and CNRS, 
             4 place Jussieu, 
             75252 Paris Cedex 05, France }
\author{L.~Julien}
\affiliation{Laboratoire Kastler Brossel, ENS, UPMC and CNRS, 
             4 place Jussieu, 
             75252 Paris Cedex 05, France }
\author{P.E.~Knowles}
\affiliation{D\'epartement de Physique, Universit\'e de Fribourg,
             1700 Fribourg, Switzerland }
\author{F.~Kottmann}
\affiliation{Institut f\"ur Teilchenphysik, ETH Z\"urich, 
             8093 Z\"urich, Switzerland }
\author{Y.-W.~Liu}
\affiliation{Physics Department, National Tsing Hua University, Hsinchu 300, 
              Taiwan }
\author{J.A.M.~Lopes}
\affiliation{Departamento de F{\'\i}sica, Universidade de Coimbra, 
             3000 Coimbra, Portugal  }
\author{C.M.B.~Monteiro}
\affiliation{Departamento de F{\'\i}sica, Universidade de Coimbra, 
             3000 Coimbra, Portugal  }
\author{F.~Mulhauser} 
\altaffiliation[Present address: ]{ University of Illinois at Urbana-Champaign,
                 Urbana, IL 61801, USA}
\affiliation{D\'epartement de Physique, Universit\'e de Fribourg,
             1700 Fribourg, Switzerland }
\author{F.~Nez}
\affiliation{Laboratoire Kastler Brossel, ENS, UPMC and CNRS, 
             4 place Jussieu, 
             75252 Paris Cedex 05, France }
\author{R.~Pohl}
\affiliation{Max-Planck-Institut f\"ur Quantenoptik, 85748 Garching, 
             Germany}
\affiliation{Paul Scherrer Institut, 5232 Villigen-PSI, Switzerland }
\author{P.~Rabinowitz}
\affiliation{Department of Chemistry, Princeton University, Princeton, NJ
             08544-1009, USA  }
\author{J.M.F.~dos~Santos}
\affiliation{Departamento de F{\'\i}sica, Universidade de Coimbra, 
             3000 Coimbra, Portugal  }
\author{L.A.~Schaller}
\affiliation{D\'epartement de Physique, Universit\'e de Fribourg,
             1700 Fribourg, Switzerland }
\author{C.~Schwob}
\affiliation{Laboratoire Kastler Brossel, ENS, UPMC and CNRS, 
             4 place Jussieu, 
             75252 Paris Cedex 05, France }
\author{D.~Taqqu}
\affiliation{Paul Scherrer Institut, 5232 Villigen-PSI, Switzerland }
\author{J.F.C.A.~Veloso}
\affiliation{Departamento de F{\'\i}sica, Universidade de Coimbra, 
             3000 Coimbra, Portugal  }
\date{\today}

\begin{abstract}
Metastable ${2S}$ muonic-hydrogen atoms undergo
collisional ${2S}$-quenching, with rates which depend strongly on
whether the $\mu p$ kinetic energy is above or below the ${2S}\to {2P}$ energy
threshold.
Above threshold, collisional ${2S} \to {2P}$ excitation
followed by fast radiative ${2P} \to {1S}$ deexcitation is allowed.
The corresponding short-lived $\mu p ({2S})$ component was measured at
0.6~hPa $\mathrm{H}_2$ room temperature gas pressure, with lifetime
$\tau_{2S}^\mathrm{short} = 165 ^{+38} _{-29}$~ns (i.e.,
$\lambda_{2S}^\mathrm{quench} = 7.9 ^{+1.8}_{-1.6} \times 10^{12} 
\, \mathrm{s}^{-1}$ at liquid-hydrogen density)
and population $\varepsilon_{2S}^\mathrm{short} = 1.70^{+0.80} _{-0.56}$\,\% 
(per $\mu p$ atom).
In addition, a value of the $\mu p$ cascade time, 
$T_\mathrm{cas}^{\mu p} = (37\pm5)$~ns, was found.
\end{abstract}

\pacs{34.50.Fa, 36.10.Dr} %

\maketitle

Muonic hydrogen ($\mu^- p$) is a simple atomic system, sensitive to
basic features of the electromagnetic and weak interactions.
Of particular interest is its metastable ${2S}$ state, long sought
after for measuring the $\mu p ({2S})$-Lamb shift, ${\cal L}_{\mu}$.
Vacuum polarization dominates  ${\cal L}_\mu$. It shifts the ${2S}$ level
by $-0.2$~eV below the ${2P}$ level~\cite{pachu96,eides01}.
A measurement of ${\cal L}_\mu$ is in progress at the Paul Scherrer
Institute (PSI), Switzerland~\cite{pohlx05}.
We report here the data analysis of a preliminary stage of this
experiment made at low $\mathrm{H}_2$ gas pressure
$p_{\mathrm{H}_2}=0.6$~hPa~\cite{ludhophd}.

Muons stopped in $\mathrm{H}_2$ gas
form highly excited $\mu p$ atoms~\cite{jense02}.
A cascade of both collisionally-induced and radiative deexcitations
leads to the ${1S}$ ground state or, with a probability
$\varepsilon_{2S}$ (few \%), to the ${2S}$ state.
The $\mu p ({1S})$ kinetic energy distribution
$E_\mathrm{kin}^{1S}$ has been measured at $p_{\mathrm{H}_2} = 0.06
\cdots 16$~hPa~\cite{pohlxphd}, and cascade simulations
show that $E_\mathrm{kin}^{1S}$ and $E_\mathrm{kin}^{2S}$ do
not differ significantly under our
conditions~\cite{jense02}.
The ${2S}$ state lifetime is, in absence of collisions,
essentially equal to the muon lifetime $\tau_\mu =2.2\:\mu$s.
In $\mathrm{H}_2$ gas, collisional ${2S}$-quenching occurs, with
different processes for kinetic energies $E_\mathrm{kin}^{2S}$ above
or below the ${2S} \to {2P}$ transition threshold which is $
(1 + m_{\mu p}/m_{\mathrm{H}_2}) |{\cal L}_\mu| \, \approx 0.3$~eV in
the lab frame:

({\it i\/})   Most $\mu p ({2S})$ atoms are formed at energies above
             this threshold~\cite{pohlxphd} where a collisional ${2S}
             \to {2P}$ Stark transition, followed by ${2P} \to {1S}$
             deexcitation with 1.9~keV $K_\alpha$ x~ray emission
            (the ${2P}$-lifetime is 8.5~ps),
            \begin{equation}
             {\mu p}(2S) +\mathrm{H}_2 \to {\mu p}(2P) + \mathrm{H}_2 
              \to {\mu p}(1S) + \mathrm{H}_2 +K_\alpha \, ,
            \label{eq:quenching}
            \end{equation}
             leads to fast ${2S}$-depletion (collisional quenching). 
             A~ lifetime $\tau_{2S}^\mathrm{short} \sim 
             100$~ns/$p_{\mathrm{H}_2}$[hPa]
             was predicted for this {\it short-lived} 
             ${2S}$-component~\cite{kodos71,carbo77,jense00}.
             This is too short to have been seen in previous searches
             for $K_\alpha$ x~rays delayed with respect to the
             $\mu p ({2S})$ formation
             time~\cite{ander77,eganx81,boeckphd}.
             In this Letter we report on the first measurement of
             $\tau_{2S}^\mathrm{short}$ and the corresponding population
             $\varepsilon_{2S}^\mathrm{short}$.

({\it ii\/}) Due to elastic collisions, a fraction of the $\mu p (2S)$
             atoms decelerates to energies below 0.3~eV where
             process~(\ref{eq:quenching}) is energetically forbidden.
	     This fraction is the {\it long-lived} ${2S}$-component.
             A recent experiment~\cite{pohlx06} showed that its
             dominant quenching process is nonradiative deexcitation
             to the ground state, with lifetime
             $\tau_{2S}^\mathrm{long} \approx 1.3 \, \mu$s at 0.6~hPa,
             and population $\varepsilon_{2S}^\mathrm{long} \approx
             1$\%.

The cascade time $T_\mathrm{cas}^{\mu p}$ is the mean delay between
$\mu p$-atom formation and final deexcitation to the ground state (when
a $\mu p$ $K$-line x~ray, other than from $\mu p (2S)$ decay,
is emitted).
The $T_\mathrm{cas}^{\mu p}$ value results from the average
of the various cascade deexcitation processes, 
and depends on $p_{\mathrm{H}_2}$.
It was calculated~\cite{jense02} but never measured for $\mu p$.

\vspace{0.1\baselineskip} 
In our experiment, muons stop in $\mathrm{H}_2$ gas containing a small
admixture of $\mathrm{N}_2$ (air), and we measure simultaneously the
three time distributions:
\vspace{-0.5\baselineskip}
\begin{itemize}
\setlength{\itemsep}{-0.2\baselineskip}
\newlength{\lbox}
\settowidth{\lbox}{$K_{\!>\!\beta}$}
  \item[--] \makebox[\lbox][l]{$K_\alpha$}, i.e., 
     $\mu p${\small ($n = 2 \to 1$)} x~rays (1.898~keV),
  \item[--] \makebox[\lbox][l]{$K_{\!>\!\beta}$}, i.e., 
     $\mu p${\small ($n > 3 \to 1$)} x~rays (2.45(2)~keV~\cite{ander84}),
  \item[--] \makebox[\lbox][l]{$\mu \mathrm{N}$}, i.e., 
     $\mu \mathrm{N}${\small ($n = 5 \to 4$)} x~rays (3.08~keV~\cite{hause98}).
\end{itemize}
\vspace{-0.3\baselineskip}
The $\mu p${\small (${n\!=\!3\!\to\!1}$)} $K_{\beta}$-line (2.249~keV)
is not well separated from $K_\alpha$ and $K_{\!>\!\beta}$.
The 0.4(1)\% air admixture in the $\mathrm{H}_2$ was
useful for calibrating x-ray energies and times.
The $\mu \mathrm{N}$ time distribution is similar to that of $\mu
p$ formation because the $\mu \mathrm{N}$ cascade time is negligibly
short ($\sim\!10^{-10}$~s)~\cite{bracc78}. 
The $\mu p$ cascade time will therefore show up as a time delay
in the $K_\alpha$ and $K_{\!>\!\beta}$ distributions compared to $\mu
\mathrm{N}$.
The signature for ${2S}$ radiative decay will be a tail in $K_\alpha$
not present in $K_{\!>\!\beta}$.
The muon transfer rates $\mu p+\mathrm{N} \to \mu \mathrm{N} + p$
are $\sim\!10^3 \, \mathrm{s}^{-1}$ for $\mu p(1S)$ 
and $\sim\!10^4 \, \mathrm{s}^{-1}$ for $\mu p(2S)$~\cite{bracc79}, 
too small to affect our results.

The experiment was performed at the recently developed low-energy
$\mu^-$ source attached to the $\pi$E5 beam line at
PSI~\cite{antog05}.  
It provides $\sim 10^3 \, \mathrm{s}^{-1} \, \mu^-$ with energies of a
few keV\@.
The muons were axially injected into a 1~m long, 20~cm bore solenoid
operated at 5~T, containing the muon entrance
detectors and the gas target (see~\cite{pohlx05}).
Two detectors, based on nanometer-thick carbon foils,
signaled the arrival of slow muons~\cite{muhlb99}.
Muons were stopped in a 20~cm long target vessel filled with
0.6~hPa $\mathrm{H}_2$ gas (temperature 290~K), and $\mu p$ atoms were
formed in a volume of $0.5 \times 1.5 \times 20~\mbox{cm}^3$.
Twenty large-area avalanche photodiodes (LAAPD), each with sensitive area
$13.5 \times 13.5~\mbox{mm}^2$, were used as x-ray
detectors~\cite{ludho05}.
Muon-decay electrons were detected by a set of plastic scintillators
and also by the LAAPDs.
More than $5\times 10^5$ events, each where an x~ray was followed by an
electron, were analyzed.

\begin{figure}
  \centerline{
  \includegraphics[width=\columnwidth]{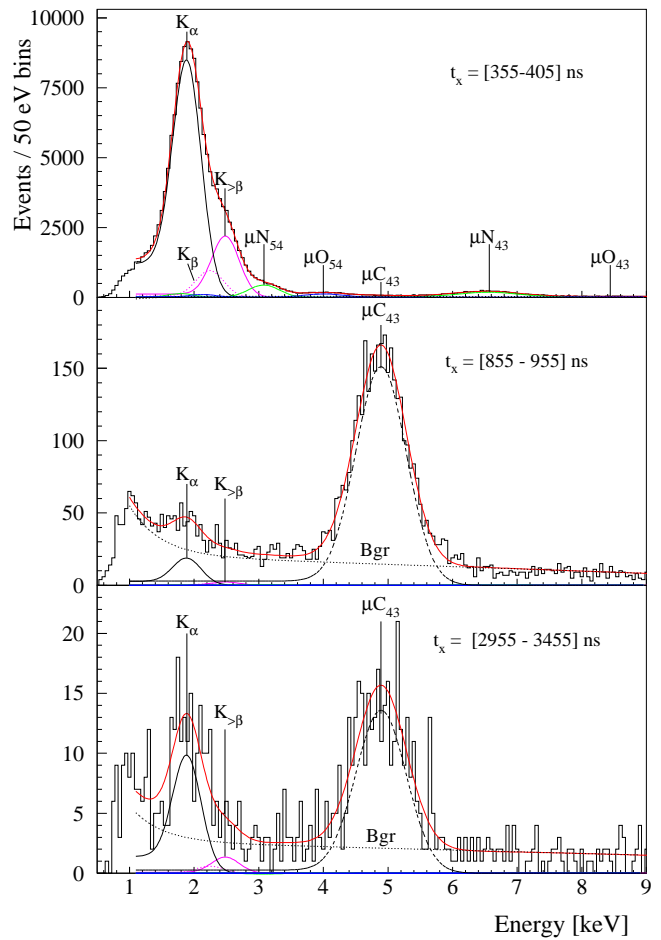}}
  \vspace*{-0.5\baselineskip}
  \caption{(color online). X-ray energy spectra (3 of 28) for
         different $t_\mathrm{x}$-intervals.  The fit function 
         is composed of the peaks for $\mu p$ $K_\alpha$,
         $K_{\beta}$, $K_{\!>\!\beta}$,
         $\mu \mathrm{N}$, $\mu \mathrm{O}$,
         $\mu \mathrm{C}$(4.89~keV),
         and a continuous background (Bgr).}
  \label{fig:x_energy}
\end{figure}

Calibration data were used for each LAAPD to deduce the energy
$E_\mathrm{x}$ and time $t_\mathrm{x}$ (relative to muon entrance)
of a measured x~ray.
Typical resolutions (full width at half maximum)
were $\Delta E_\mathrm{x} /E_\mathrm{x} \approx 25$\% 
and $\Delta t_\mathrm{x} \approx 35$~ns for 2-keV x~rays.
Most $\mu p$ x~rays were found in the time interval $300 \le
t_\mathrm{x} \le 600$~ns, corresponding to the widely-distributed
muon slowing-down times.

The $K_\alpha$, $K_{\!>\!\beta}$, and $\mu \mathrm{N}$ time distributions
were determined from a fit of the $E_\mathrm{x}$ spectra for different
$t_\mathrm{x}$ intervals.
The useful $t_\mathrm{x}$ range ($0.2 \cdots 6.5 \, \mu$s) was divided
into 28 intervals (50, 100, and 500~ns wide).  For each interval an
$E_\mathrm{x}$ spectrum was produced.  Three typical spectra are shown
in Fig.~\ref{fig:x_energy}, one at times of $\mu p$ formation and
deexcitation (top), and two at later times.
The function fit to each spectrum is composed of several x-ray lines
and a continuous background.
Each line is the sum of a Gaussian peak and a tail towards lower
energies (an LAAPD characteristic), with energy-dependent
weights~\cite{ludhophd}.

The lines fitted in Fig.~\ref{fig:x_energy} correspond to x~rays from
$\mu p$ ($K_\alpha$, $K_{\beta}$, and $K_{\!>\!\beta}$), $\mu \mathrm{N}$ 
(main transitions at 1.67, 3.08, and 6.65~keV~\cite{hause98}), 
$\mu \mathrm{O}$ (2.19, 4.02, and 8.69~keV), and $\mu \mathrm{C}$
(4.89~keV).
The intensities for the $K_\alpha$, $K_{\!>\!\beta}$, 3.08, 4.02, 4.89, and
6.65~keV lines were free parameters, whereas the relative intensities
of the other lines as well as the positions and widths of all lines
were fixed by requiring global consistency for all
data~\cite{ludhophd} and considering known yields~\cite{kirch99}.
The late-time 2-keV peak (see Fig.~\ref{fig:x_energy}, bottom) is
due to $\mu p$ x~rays from {\it second-muon} stops, i.e., muons
entering the target at random times shortly after a {\it first-muon}
which defined $t_\mathrm{x} = 0$.
The $\mu \mathrm{C}_{4 \to 3}$ (4.89~keV) line at late times arises from
$\mu p$ atoms drifting to the polypropylene foils in front of the
LAAPDs where muon transfer to C atoms occurred.
The continuous background is due to x~rays with energy $>\!10$~keV,
e.g., $\mu \mathrm{C}$ $K$ and $L$ line transitions, whose deposited
charge was not fully amplified by the LAAPDs.
It was well modeled by a sum of an exponential and a linear
function, with two amplitudes as free parameters.  
The full details of the extensive background studies 
are found in~\cite{ludhophd}.

\begin{figure}
  \centerline{
  \includegraphics[width=\columnwidth]{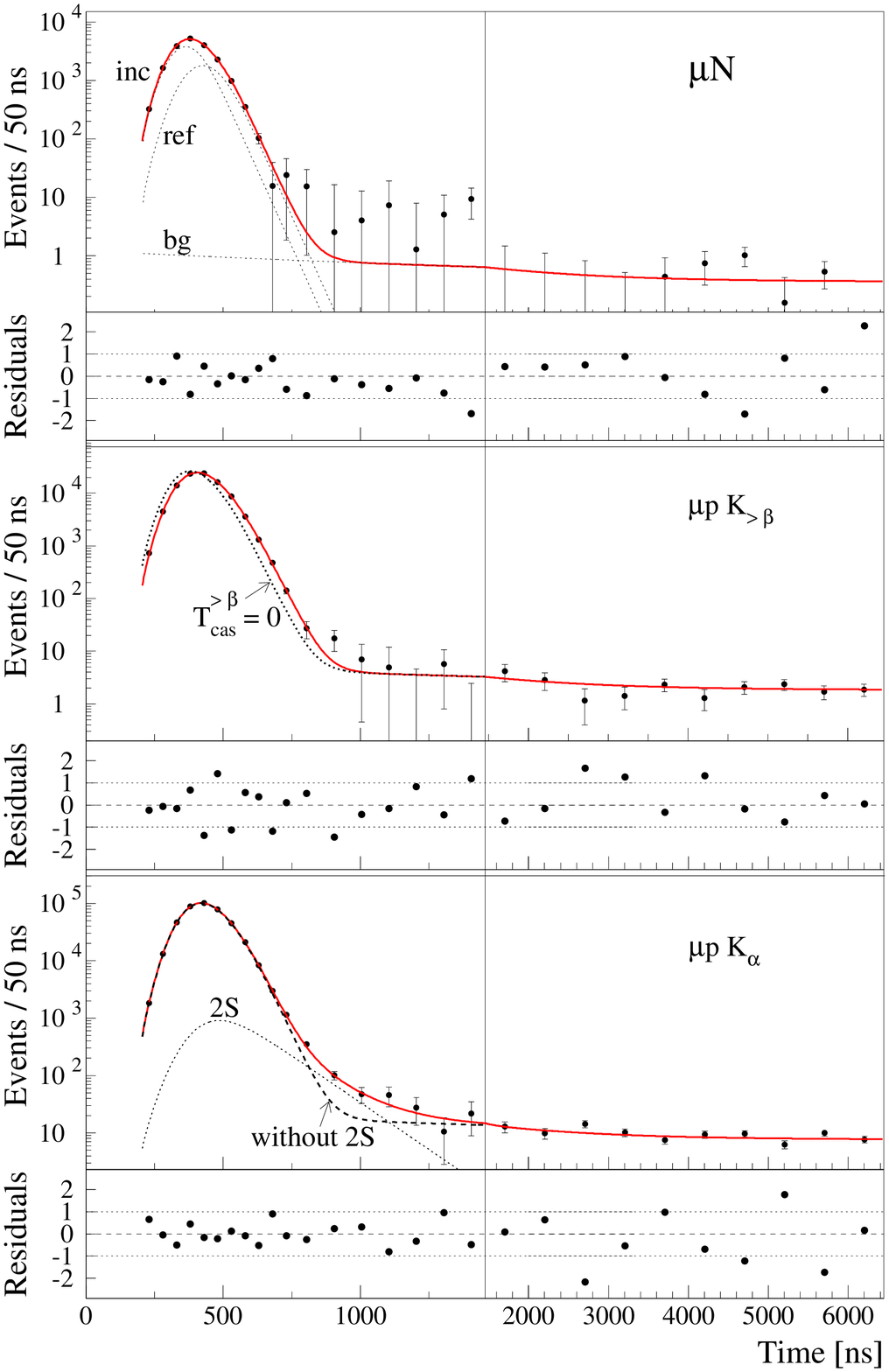}}
  \vspace*{-2.0\baselineskip}
  \caption{
    Fit functions (solid line) and residuals (normalized by the
    errors) for fits to the x-ray time spectra for the lines $\mu
    \mathrm{N}$(3.08~keV) (top), $\mu p - K_{\!>\!\beta}$ (middle),
    and $K_\alpha$ (bottom).  Each point results from a fit of the
    x-ray energy spectrum for the corresponding
    $t_\mathrm{x}$-interval (50, 100, and 500~ns wide).
     The dotted lines in the $\mu \mathrm{N}$ spectrum are 
     the functions for incoming (inc), reflected (ref), and
     {\it second} (bg) muons.  In the $K_{\!>\!\beta}$ spectrum the
     dotted line is the fit function without cascade time.
     In the $K_\alpha$ spectrum the ${2S}$-tail, i.e., the contribution
     of the short-lived $\mu p ({2S})$ state, is shown as dotted line. 
     The dashed line is the fit function minus the ${2S}$-tail.}
  \label{fig:fit}
\end{figure}

The fitted intensities (with statistical errors) of
the 28~energy spectra are shown in Fig.~\ref{fig:fit} as a function of
$t_\mathrm{x}$, for the three lines $\mu \mathrm{N}$(3.08~keV), $\mu p$
$K_{\!>\!\beta}$, and $K_\alpha$.
The late-time events are caused by {\it second-muon} stops.
The $t_\mathrm{x}$ spectra from different LAAPD positions 
show~\cite{ludhophd} that
for muons not stopped during their first pass through the gas
target, a considerable fraction was reflected at the gold-plated back
side of the vessel and stopped during the return pass.
Consequently, the $\mu \mathrm{N}$ spectrum (showing the muon
stop-time distribution) was represented by a sum of two functions,
each one the convolution of a Gaussian with an exponential
(Fig.~\ref{fig:fit}, top).
The simultaneous fit of the three time spectra, using the common muon
stop-time distribution, gives values for the intensities $A_{\mu
\mathrm{N}}$, $A_{>\!\beta}$, and $A_\alpha$.

The $K_{\!>\!\beta}$ and $K_\alpha$ spectra are delayed and slightly
broadened with respect to $\mu \mathrm{N}$ due to the $\mu p$ cascade
time.
It is not possible to extract the precise shape of the cascade time
distribution from the data, but calculations~\cite{jense02} indicate
that the $K_{\!>\!\beta}$ cascade has an approximately exponential time
distribution, whereas the $K_\alpha$ cascade has the same asymptotic
behaviour but includes a ``build-up'' character at earlier times.
The $\mu \mathrm{N}$ fit function was therefore convoluted with an
exponential (parameter $\tau_\mathrm{cas}$) to obtain the $K_{\!>\!\beta}$
function and further convoluted with a Gaussian (parameter
$\sigma_\mathrm{cas}^\alpha$) for the $K_\alpha$ function.
In addition, free time offsets $\Delta T_{\!>\!\beta}$ and $\Delta
T_\alpha$ (for $K_{\!>\!\beta}$ and $K_\alpha$, with respect to $\mu
\mathrm{N}$) were introduced.
The contribution of the short-lived $\mu p ({2S})$ state was
considered by adding to the $K_\alpha$ function a convolution of the
$K_{\!>\!\beta}$ distribution with an exponential (parameter
$\tilde{\tau}_{2S}^\mathrm{short}$), and a relative population
$\tilde{\varepsilon}_{2S}^\mathrm{short}$ (normalized for the fit to
$A_\alpha$).

A simultaneous fit of the three time spectra
was performed, resulting in $\chi^2 = 57.5$ for 67~degrees of freedom. 
The fit functions and residuals are shown in Fig.~\ref{fig:fit}.
The scatter of the residuals confirms that no relevant systematic
deviations exist between the data and the model.
The resulting cascade time slope is $\tau_\mathrm{cas} = (26\pm2)$~ns.
The time shift $\Delta T_{\!>\!\beta} = (0\pm5)$~ns is consistent with
zero, as expected for the $K_{\!>\!\beta}$ cascade, whereas $\Delta
T_\alpha = (13\pm5)$~ns and $\sigma_\mathrm{cas}^\alpha = (15\pm2)$~ns
approximate a $K_\alpha$ cascade time distribution which deviates from
an exponential in the first $\sim\! 20$~ns.
The deduced mean cascade times $T_\mathrm{cas}^{>\!\beta} =
\tau_\mathrm{cas} + \Delta T_{\!>\!\beta} = (26\pm5)$~ns and
$T_\mathrm{cas}^\alpha = \tau_\mathrm{cas} + \Delta T_\alpha = (39\pm5)$~ns,
weighted by the corresponding $K$-line yields, result in 
a mean $\mu p$ cascade time $T_\mathrm{cas}^{\mu p} = (37\pm5)$~ns.
(The relative $K$-yields at $p_{\mathrm{H}_2} = 0.6$~hPa
were deduced from~\cite{ander84} as $Y_\alpha = 0.821(12)$,
$Y_{\beta} = 0.061(9)$, and $Y_{\!>\!\beta} = 0.118(11)$,
and the $K_{\beta}$ cascade time was assumed to be equal to
$T_\mathrm{cas}^{>\!\beta}$.)
$T_\mathrm{cas}^{\mu p}$ depends only weakly on the fit
function details because it essentially reproduces the
center-of-gravity shifts of the $K_\alpha$ and $K_{\!>\!\beta}$
distributions relative to $\mu \mathrm{N}$.

\begin{figure}
  \centerline{
  \includegraphics[width=0.65\columnwidth]{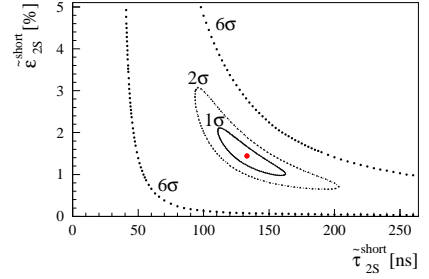}}
  \vspace*{-1.2\baselineskip}
  \caption{Relative population
       $\tilde{\varepsilon}_{2S}^\mathrm{short}$ (normalized to
       $A_\alpha$) of the short-lived $\mu p ({2S})$ component versus
       its lifetime $\tilde{\tau}_{2S}^\mathrm{short}$ (without
       systematic corrections). 
       The dot represents the best fit.}
  \label{fig:A2S_tau2s}
\end{figure}

The fit results for the ${2S}$-tail are 
$\tilde{\tau}_{2S}^\mathrm{short} = 133 ^{+29} _{-22}$~ns and
$\tilde{\varepsilon}_{2S}^\mathrm{short} = 1.44 ^{+0.67}_{-0.46}$\%. 
As shown in Fig.~\ref{fig:A2S_tau2s}, the absence of the 
${2S}$-tail, corresponding to $\tilde{\varepsilon}_{2S}^\mathrm{short}=0$, 
is excluded by $6 \sigma$.
As a test, the $K_{\!>\!\beta}$ spectrum was also fit with a
``${2S}$''-tail.  Its amplitude  $\varepsilon_{>\!\beta}$ 
(normalized to $A_{>\!\beta}$) is compatible with zero 
(within $0.7 \sigma$) for any $\tilde{\tau}_{2S}^\mathrm{short}$, 
as expected.
We conclude that our fit function correctly reproduces the muon
stop, cascade, and {\it second-muon} time distributions.
The difference between $\tilde{\varepsilon}_{2S}^\mathrm{short}$
and $\varepsilon_{>\!\beta}$ is 
$\tilde{\varepsilon}_{2S}^\mathrm{short} - \varepsilon_{>\!\beta}
 = (1.36 \pm 0.28)$\%\; at $\tilde{\tau}_{2S}^\mathrm{short} = 133$~ns.
A zero value of this difference is excluded by $\ge\! 4.4\, \sigma$
for any $\tilde{\tau}_{2S}^\mathrm{short}$,
confirming that the tail in the $K_\alpha$ spectrum can only come from 
${2S}$-quenching.

The measured $\tilde{\tau}_{2S}^\mathrm{short}$ and
$\tilde{\varepsilon}_{2S}^\mathrm{short}$ values have to be corrected
for several effects. 
The necessary corrections were deduced from a Monte Carlo simulation
of the experiment, based on the known distribution of
$E_\mathrm{kin}^{2S}$~\cite{pohlxphd} and on calculated cross
sections~\cite{jense00} for process (\ref{eq:quenching}) and elastic
collisions.
It was found that ({\it i\/}) some $\mu p ({2S})$ atoms reach the target
walls before being quenched; 
({\it ii\/}) the solid angle for $K_\alpha$
detection varies with time due to the $\mu p ({2S})$ motion; 
({\it iii\/}) because of collisions,
the $E_\mathrm{kin}^{2S}$ distribution depends on time,
and hence so do the mean cross sections and $\mu p$ velocities.
Those effects modify the ${2S}$ tail shape at the late times
where the experiment is most sensitive.
The resulting correction factors are $1.24 \pm 0.07$ for
$\tilde{\tau}_{2S}^\mathrm{short}$ and $1.34 \pm 0.09$ for
$\tilde{\varepsilon}_{2S}^\mathrm{short}$.
In addition, $\tilde{\varepsilon}_{2S}^\mathrm{short}$ has to be
multiplied by $(1 - \tau_{2S}^\mathrm{short} / \tau_\mu)^{-1}$ to
account for muon decay and by $Y_\alpha /(1 +
\varepsilon_{2S}^\mathrm{long})$ to normalize to all $\mu p$ atoms.
The final result for the radiative lifetime and population of the
short-lived $\mu p ({2S})$ component at 0.6~hPa (temperature 290~K) is
$\tau_{2S}^\mathrm{short} = 165 ^{+38} _{-29}$~ns and
$\varepsilon_{2S}^\mathrm{short} = 1.70 ^{+0.80} _{-0.56}$\,\%.

\vspace{0.2\baselineskip}
The $\mu p$ cascade time and the ${2S}$-tail have been
disentangled from the stop-time distribution for the first time. 
This has been made possible by
the high statistics, the low gas pressure, and the good stop-time
resolution.
The measured $\mu p$ cascade time $T_\mathrm{cas}^{\mu p} =(37\pm5)$~ns 
is less than half the $\approx \!90$~ns value predicted for 0.6~hPa
by cascade calculations~\cite{jense02}.
The measured value may be slightly affected by
a possible difference in the $\mu^-$
atomic capture times of $\mathrm{N}_2$ and $\mathrm{H}_2$ 
predicted by some models~\cite{cohen04}:
the muon energy from which capture occurs is expected to increase with
$Z$~\cite{cohen02}, so $\mu \mathrm{N}$ atoms can form earlier than
$\mu p$ atoms during the muon stopping, an effect which
would result in an even lower measured $T_\mathrm{cas}^{\mu p}$ value.
A result which does not depend on such effects
is the measured difference 
$T_\mathrm{cas}^\alpha - T_\mathrm{cas}^{>\!\beta} = (13\pm4)$~ns, 
also significantly smaller than the calculated value of $\approx \!25$~ns. 
The calculated cascade times may be too long since
Coulomb deexcitations, which dominate the cascade at high
$n$-levels even at low $p_{\mathrm{H}_2}$~\cite{jense02}, may be
accompanied by {\it simul\-taneous} Auger transitions, an effect not yet
considered.

We have compared the measured radiative ${2S}$-lifetime
with the results of calculations.
Since the existing calculated cross sections 
neglect molecular effects we allowed for
the extreme cases where the $\mathrm{H}_2$ molecule is taken as two separate
H~atoms or as a single atom. 
Analysing the simulated data in the time region where the
experiment is sensitive, we obtained a value of ($178 \pm 30$)~ns
which agrees well with the experimental result
$\tau_{2S}^\mathrm{short} = 165 ^{+38}_{-29}$~ns.
This confirms the
validity of the cross sections calculated for 
$\mu p ({2S}) + \mathrm{H} \to \mu p ({2P}) + \mathrm{H}$\;\; 
Stark transitions~\cite{carbo77,jense00}.
The quenching rate for the short-lived ${2S}$-component, when
normalized to liquid-hydrogen atom density (LHD, $4.25 \times 10^{22}$
atoms/cm$^3$), is $\lambda_{2S}^\mathrm{quench}(\mathrm{LHD}) = 
7.9 ^{+1.8} _{-1.6} \times 10^{12} \, \mathrm{s}^{-1}$, 
$\sim 20$~times faster than for the long-lived one~\cite{pohlx06}.

The sum of the measured relative populations
$\varepsilon_{2S}^\mathrm{short} = 1.70 ^{+0.80} _{-0.56}$\,\% and
$\varepsilon_{2S}^\mathrm{long} \approx 1$\% (extrapolated to
0.6~hPa, from~\cite{pohlx06}) is ($2.7 \pm 0.8$)\%,
in agreement with the total ${2S}$-population $\varepsilon_{2S} =
(2.49 \pm 0.17)$\% at 0.6~hPa deduced
directly from the measured $\mu p$ $K$-line yields~\cite{ander84}.

We conclude that there is now good understanding
of both the short- and long-lived $\mu p ({2S})$ dynamics.

\begin{acknowledgments}
We thank L.M.~Simons and B.~Leoni for setting up the cyclotron trap,
O.~Huot and Z.~Hochman for technical support, and T.S.~Jensen
for fruitful discussions.  We also thank the PSI accelerator division,
the Hallendienst, and the workshops at PSI, MPQ, and Fribourg
for their valuable help.
This work was supported by the Swiss National Science Foundation, the
French-Swiss program ``Germaine de~Sta{\"{e}}l'' (PAI n$^\circ$07819NH), 
the BQR de
l'UFR de physique fondamentale et appliqu\'ee de l'Universit\'e Paris 6,
the Portuguese FCT and FEDER under project POCTI/FNU/41720/2001,
and the US Department of Energy.

\end{acknowledgments}

\vspace{-0.5\baselineskip}
\end{document}